\documentclass[12pt]{article}

\usepackage{hyperref}

\usepackage{tikz}

\usepackage{epsf}
\usepackage{amsmath,amssymb}
\usepackage{latexsym}
\usepackage{graphicx,color}
\usepackage{wrapfig}
\usepackage{latexsym}
\usepackage{graphics}

\usepackage{color}
\usepackage{delarray,color}
\usepackage{fancybox}  

\newcommand {\beq}{\begin{align}}
\newcommand {\eeq}{\end{align}}

\newcommand{\be}{\begin{equation}}
\newcommand{\ba}{\begin{align}}
\newcommand{\ea}{\end{align}}
\newcommand{\ee}{\end{equation}}

\newcommand{\beqa}{\begin{align}}
\newcommand{\eeqa}{\end{align}}

\newcommand{\unit}{\hbox to 3.8pt{\hskip1.3pt \vrule height 7.4pt
    width .4pt \hskip.7pt \vrule height 7.85pt width .4pt \kern-2.4pt
    \hrulefill \kern-3pt \raise 3.7pt\hbox{\char'40}}}

\def\matt[#1,#2,#3,#4]{\left(%
\begin{array}{cc} #1 & #2 \\ #3 & #4 \end{array} \right)}

\usepackage{tabularx}

\catcode`\@=11
\@addtoreset{equation}{section}

\catcode`@=12
\relax

\begin{document}

\begin{titlepage}

\setcounter{page}{0}

\renewcommand{\thefootnote}{\fnsymbol{footnote}}

\begin{flushright}
YITP-25-95
\end{flushright}

\vskip 1.35cm

\begin{center}
{\Large \bf 
Stretched Horizon Dissipation and the Fate of Echoes
}

\vskip 1.2cm 

{\normalsize
Seiji~Terashima\footnote{terasima(at)yukawa.kyoto-u.ac.jp}
}

\vskip 0.8cm

{ \it
Center for Gravitational Physics and Quantum Information,

Yukawa Institute for Theoretical Physics, Kyoto University, Kyoto 606-8502, Japan
}

\end{center}

\vspace{12mm}

\centerline{{\bf Abstract}}

We examine the dissipation of infalling particles near black holes under the assumption that the equivalence principle breaks down at the stretched horizon. This violation allows partial reflection of particles just outside the event horizon, as suggested by models such as the fuzzball and firewall proposals. We show that scattering with blue-shifted Hawking radiation leads to moderate dissipation when the particle energy is comparable to the Hawking temperature. The decay rate is independent of the Planck mass, and therefore gravitational wave echoes-arising from such partial reflection-may survive despite the presence of quantum gravitational effects. Our result is universal and does not rely on the detailed microphysics of the horizon.


\end{titlepage}
\newpage

\tableofcontents
\vskip 1.2cm

\section{Introduction and summary}

What happens when a particle falls into a black hole? 
According to (classical) general relativity, it crosses the event horizon and enters the interior, never to return. 
Due to semiclassical effects, the particle may very slowly escape from the black hole as Hawking radiation \cite{Hawking:1975vcx}. 
However, this picture has not been observed or experimentally verified, and it is not necessarily correct in (non-perturbative) quantum gravity.
In fact, if the fuzzball proposal \cite{Mathur:2005zp} \cite{Mathur:2009hf} (also known as the firewall proposal \cite{Almheiri:2012rt}), which was suggested by string theory and quantum information theoretical arguments, is correct, then the equivalence principle can no longer guarantee that nothing happens when crossing the horizon.
In particular, because the degrees of freedom of a black hole are finite, simple models like the brick wall model \cite{tHooft:1984kcu}, which implement this by using Dirichlet boundary conditions, posit that the degrees of freedom only extend to a surface called the stretched horizon \cite{tHooft:1984kcu} \cite{Susskind:1993if}, located a Planck length away from the event horizon.\footnote{
The brick wall model was introduced in \cite{tHooft:1984kcu} but was not regarded as an appropriate model because of the violation of the equivalence principle. Here, we study the possibility of the violation of it. Note that it is argued \cite{Iizuka:2013kma} \cite{Terashima:2020uqu} \cite{Sugishita:2023wjm} that a structure like the brick wall will appear in the black hole in AdS/CFT correspondence \cite{Maldacena:1997re}.
}
In such models, incoming particles are totally reflected at the stretched horizon.

While this is purely a theoretical consideration and might seem unobservable in practice, the observations of gravitational waves (GWs) from black hole mergers may provide evidence of the non-perturbative quantum-gravitational effects near the horizon \cite{Cardoso:2016rao} \cite{Cardoso:2016oxy}  \cite{Price:2017cjr}; see also \cite{Cardoso:2019rvt, Abedi:2020ujo} for reviews.
If the effect causes GWs to (partially) reflect at a location the Planckian length away from the horizon, a cavity between the reflective boundary and the angular momentum potential leads to the emission of GW echoes in the time interval of milliseconds or so for the black hole mass of $\sim {\cal O}(10) M_{\odot}$.
If the GW echoes following the main signal are observed, it may be supporting evidence of the existence of the stretched horizon.
Therefore, the issue of near-horizon geometry of black holes is of great importance, as it could be confirmed through actual observations.\footnote{
A precise calculation of the reflectivity would in principle require a full theory of quantum gravity, and is therefore challenging. However, various models have been proposed, and the reflectivity has been computed within those frameworks \cite{Mark:2017dnq} \cite{Bueno:2017hyj} \cite{Oshita:2018fqu}
\cite{Wang:2019rcf}
\cite{Oshita:2019sat}
\cite{Cardoso:2019apo}
\cite{Deppe:2024qrk}
\cite{Maggio:2020jml}
\cite{Dong:2020odp}
\cite{Chakravarti:2021clm}
\cite{Rosato:2025byu}.
}
If we consider the simple model described above as an approximation of the actual situation, it raises a question such as the one described below.
The model suggests that the black hole does not absorb anything and essentially loses its ``blackness.''
An important fact is that incoming particles experience (infinite) blue shift as they approach the horizon.
As a result, even in situations where gravitational (or other) interactions are small, they are significant near the horizon.
Since black holes emit Hawking radiation, scattering with this radiation cannot be ignored. 
Thus, particles decay and fragment into lighter particles. 
Then, one may wonder whether the incoming GWs (or gravitons and other particles) completely dissipate into blue-shifted Hawking radiation near the horizon radius.
If the dissipation is significant, one could not expect coherent and observable GW echoes.
If it turns out that the scattering cross-section grows with the Planck mass, then nearly all incoming particles would end up scattering and fragmenting into lighter particles.
This process would eventually thermalize and be viewed as Hawking radiation in semiclassical approximations. 
GW echoes would not appear in this case due to the strong dissipation.
Conversely, if the scattering cross-section is proportional to a negative power of the Planck mass, the dissipation would be negligible. 
Therefore, it is crucial to estimate the significance of dissipation effects of the scattering by the Hawking radiation.
%

In this work, we estimate the dissipation effect from a microscopic perspective using known physics: Einstein gravity, the Standard Model, and higher derivative corrections suppressed by the Planck scale. Importantly, we find that the decay rate of infalling particles depends only on the ratio of their frequency to the Hawking temperature, and is independent of the Planck mass.\footnote{Since we qualitatively examine regions not far from the black hole, the choice of asymptotic flatness or asymptotic AdS does not matter.} 

Our main result is that modes with frequency comparable to the Hawking temperature may partially reflect at the stretched horizon, while experiencing only moderate dissipation. This supports the possibility of observing GW echoes even after accounting for interactions with blue-shifted Hawking radiation. 
%
%
%
%
%
%
Our result is universal in the sense that it is independent of the details of the (yet unknown) quantum gravity theory as long as our perturbative treatment of quantum gravity effects is valid.
In fact, the effect considered in this argument occurs before the particle enters the stretched horizon. Therefore, it does not rely on any assumptions about (non-standard) physics on the stretched horizon or inside the black hole. (Strictly speaking, we approximate the effect of the strongly coupled region near the stretched horizon by the leading-order scattering process.)\footnote{
The superradiant instability discussed in~\cite{Nakano:2017fvh, Oshita:2020dox} is avoided in our setup, as the dissipation due to scattering with Hawking radiation provides an effective damping mechanism near the stretched horizon.
}
We also discuss the consistency with the Boltzmann reflectivity, which is a theoretical model of the reflectivity of a (dissipative) quantum black hole \cite{Oshita:2018fqu}.

Throughout the manuscript, $G_N$ is the gravitational constant, and we take the natural unit $c=\hbar=1$. 
In this unit, the Planckian length $l_P$ is related to $G_N$ via $G_N = l_P^2$.

\section{Particle falling into the stretched horizon}

Let us consider the Schwarzschild black hole of mass $M$, whose line element takes the form of
\begin{align}
    ds^2=-\left( 1-\frac{r_h}{r} \right) dt^2+\left( 1-\frac{r_h}{r} \right)^{-1} dr^2 +r^2 d \Omega^2, 
\end{align}
where $r_h=2 G_N M$, $t$ is the Schwarzschild time, $r$ is the areal coordinate, and $d\Omega^2$ is the line element on the two-sphere. 
Here, we consider four dimensional gravity and the cosmological constant is zero.
We introduce a coordinate $x$, which represents the proper radial distance from the horizon, as
\begin{align}
    x=\int_{r_h}^{r} \frac{1}{\sqrt{1-\frac{r_h}{r'}}} dr'
    =r_h \int_{1}^{r/r_h} \frac{r'}{\sqrt{r'-1}} dr'
    =\frac{2 \sqrt{r_h (r-r_h)} (r+2 r_h)}{3 r_h} .
\end{align}
Near the horizon, we have $x \simeq 2 \sqrt{r_h(r-r_h)}$
and the metric can be approximated as the Rindler (like) space:
\begin{align}
    ds^2=-\frac{1}{4 (r_h)^2} x^2 dt^2+dx^2 +(r_h)^2 d \Omega^2.
    \label{Rm}
\end{align}
We introduce the stretched horizon or ``brick wall'' at $r=r_b$ where
\begin{align}
    r_b=r_h+\frac{(l_p)^2}{4 r_h},
\end{align}
and $G_N = (l_p)^2$. This is equivalent to placing it at $x =l_p$.

Let us consider a wave packet of a test scalar field.
%
Here, we consider the test scalar field for notational simplicity.
We can treat other fields, including the electromagnetic field or metric perturbations, in the same way.
In the following, 
we assume $r_h \gg l_p$.
We also assume that the typical frequency of the infalling wave packet is ${\cal O} (1/l_p)$.
We also assume that the mass of the test scalar field is small ($m^2 \ll 1/(r_h)^2$), 
which implies that the mass term is negligible for our purpose.\footnote{The assumption for the interaction terms will be stated later.
Note that even if $1/(l_p)^2 \gg m^2 \gg 1/(r_h)^2$, the mass term will be negligible for the computations in this paper because of the blueshift near the horizon. }

In classical theory, nothing special happens when the wave packet approaches the (stretched) horizon, because we can consider the locally flat coordinate by the diffeomorphism. 
If we assume that the wave packet is reflected at the brick wall with the Dirichlet boundary condition,
then it leads to the emission of {\it echoes} of the waves. The time scale of the echoes is $\Delta t \equiv t_1-t_2$, where $t_1$ and $t_2$ are
the Schwarzschild time when the incoming and outgoing wave packets pass the potential barrier at $r=r_0 \sim (3/2) r_h$.
The echo time $\Delta t$ is roughly estimated with $\Delta t \sim r_h \log (r_h/l_p)$. 
Note that the wave packet does not decay in this process in classical theory
with the brick wall.

However, in the presence of quantum gravity effects, 
something dramatic may happen even before the incoming signal reaches the stretched horizon, as shown below.
The wave packet will be blue-shifted near the horizon,
and the typical frequency of the wave packet becomes comparable to the Planck mass $m_p=(l_p)^{-1}$.
Then, the gravitational interaction becomes strong, and the quantum effect can not be negligible.\footnote{
The discussions below can be applied  
even for the Rindler patch of the Minkowski space, instead of the black hole background.
We expect that the quantum gravity effect can be neglected for the wave packet in the Minkowski space.
However,
we need to consider the usual vacuum in the Minkowski space.
For the wave packet in the black hole case, 
the Rindler vacuum, which is the vacuum of the Rindler patch,
should be taken, thus these two cases are different.
(More precisely, we need to consider the particles of the Hawking radiations around the vacuum near the horizon, but it is not the thermal state.)
Here, the black hole means the black hole formed by a gravitational collapse, which is the single-sided black hole in the AdS/CFT.
For the eternal black hole, we usually consider
the Hartle-Hawking vacuum, at least, for the thermofield double state in the AdS/CFT.
This will correspond to the usual vacuum of the Minkowski space for the Rindler approximation and then the wave packet will go through the horizon like the classical case.
}
Furthermore, if we assume the brick wall-like structure,
this breaks the equivalence principle or the diffeomorphism invariance near the horizon.
Thus, the quantum gravity effect may change the behavior of the wave packet considerably.

Note that such effects are significant only at the near-horizon region where we can approximate the metric as \eqref{Rm}
because the region far from the horizon can be approximately a flat space for the particle with energy much less than $m_p$.

\subsection{Decay rate}

In the following, we estimate how much a particle falling into a black hole decays before reaching the stretched horizon. As we will see later, the decay itself is negligibly small. We then proceed to estimate the effect of scattering with Hawking radiation. Since this can be inferred from the decay rate at finite temperature, we compute the latter.

First, we consider the decay rate of the wave packet
for a short time interval.
The wave packet propagates along a null geodesic during the interval from $x=\bar{x}+\delta x$ to $x=\bar{x}$.
We will approximate the geometry by the local flat space $ds^2= -d \tilde{t}^2+dx^2 +\cdots$, where $\tilde{t}=\frac{\bar{x}}{2 r_h} t+\tilde{t}_0$ and the constant $\tilde{t}_0$ is chosen such that $\tilde{t}=-x$ is the null geodesics. With this setup, let us use the usual decay rate formula for the Minkowski space in the inertial frame.\footnote{Later, we will examine the validity of this approximation.}
Below, we will use $\tilde{t}=-x$ as an affine parameter since the wave packet propagates on the null geodesics with $\tilde{t} = -x$.
Now it is important to notice that the frequency of the wave packet $\omega (\bar{x})$ at $x=\bar{x}$ is 
\begin{align}
    \omega(\bar{x}) \simeq \frac{2 r_h}{x} \omega_0,
\end{align}
where $\omega_0$ is the frequency at  $x=2 r_h$ if we assume the Rindler space \eqref{Rm} by the blue shift effect.
For the Schwarzschild metric, we can show
$\omega_0 \simeq 1.58 \, \omega_{asym}$ 
where $\omega_{asym}$ is the frequency in the asymptotic flat region.

The decay rate of a scalar particle with energy $\omega$
to $n_F$ particles will be given by 
\begin{align}
    \frac{d \Gamma}{d \tilde{t}} = \frac{1}{2 \omega} 
    \prod_{f=1}^{n_F} 
    \left( \frac{d^3 p_f}{(2 \pi)^3} \frac{1}{2 \omega_f}\right) (2 \pi)^4 \delta^4 (P-\sum_{f=1}^{n_F} P_f) |{\cal A}|^2,
    \label{dr}
\end{align}
where ${\cal A}$ is the amplitude for the decay mode, $P$ is the four-momentum of the original particle and 
$P_f$ is the four-momentum of the $f$-th particles after the decay whose components are $(\omega_f,{\bf p}_f)$ where $\omega_f=\sqrt{ ({\bf p}_f)^2}$. 
This formula is Lorentz invariant, but not diffeomorphism invariant.

Here, we consider the decay of the particle in the near horizon region, where the particle may be effectively regarded as a massless scalar by the blue shift even if we consider a massive scalar.  
For a massless scalar particle, the decay can not be allowed by the kinematical reason. Indeed, the allowed final states will be only the fragmentation of the original particle $(\omega, {\bf p})$ to $n_F$ particles with
\begin{align}
    (\omega_f, {\bf p}_f) =(\alpha_f \, \omega, \alpha_f  {\bf p} ),
    \label{frag}
\end{align}
where $\alpha_f \geq 0$ and $\sum_{f=1}^{n_F} \alpha_f=1$.

\begin{figure}[t]
\centering
\includegraphics[width=0.9\linewidth]{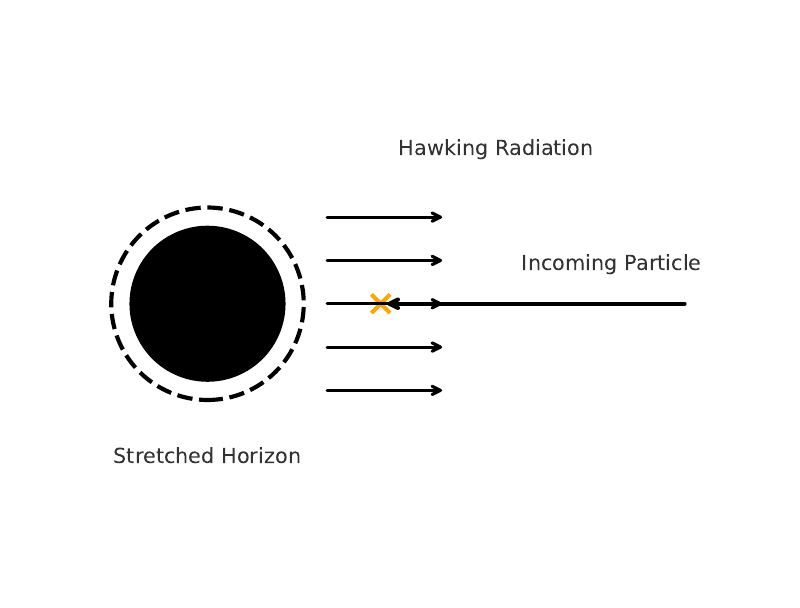}
\caption{
Schematic picture of the thermal Hawking radiation.
}
\label{fig_schematic}
\end{figure}

Thus, the ``decay" will be caused by the scattering of the particle and the Hawking radiations with the temperature
\begin{align}
    T_H=\frac{1}{4 \pi r_h}.
\end{align}
 (see Figure \ref{fig_schematic}).
We will consider the finite temperature environment instead of the outgoing Hawking radiations.   
This can be justified because we only consider the decay of the initial particle and the inverse scattering process are not large for our case.
Thus, we assume that the near horizon region is considered the (local) thermal state with the effective temperature 
\begin{align}
    T_{eff} = \frac{1}{4 \pi r_h} \sqrt{\frac{r}{r-r_h}} \simeq \frac{1}{2 \pi x},
\end{align}
although the whole spacetime is not thermal.

\begin{figure}

\begin{center}
\begin{tikzpicture}[scale=1.3, every node/.style={font=\small},
    incoming/.style={->, thick, shorten <=2pt, shorten >=2pt, blue},
    outgoing/.style={->, thick, red, shorten >=2pt},
    curvedarrow/.style={->, thick, blue!50!black}
  ]

  \draw[incoming] (-5,0) -- (-1,0) node[midway, below left] {$\omega$};
  \draw[incoming, red] (0,1) -- (-1,0) node[midway, below right] {$|\omega_{n_F}|$};

  \filldraw[black] (-1,0) circle (1.2pt);
  \node at (-1.6,-0.3) {collision};
  \node at (-4,-2.2) {Before collision};

  \draw[outgoing] (4,0) -- ++(-100:1.5) node[below right] {$\omega_1$};
  \draw[outgoing] (4,0) -- ++(-30:2) node[below right] {$\omega_2$};
  \draw[outgoing] (4,0) -- ++(30:2) node[above right] {$\omega_3$};
   \draw[outgoing] (4,0) -- ++(60:1.5) node[above right] {$\cdots$};
  \draw[outgoing] (4,0) -- ++(120:2) node[above right] {$\omega_{n_F-1}$};

  \filldraw[black] (4,0) circle (1.2pt);
  \node at (4.3,-0.3) {scattering};
  \node at (4,-2.2) {After collision};


\end{tikzpicture}
\end{center}
\caption{
Parametrization of the particles before and after the scattering.
Even the Hawking radiation  with a negative $\omega_{n_F}$ which collides with the particle is regarded as "final" particles in \eqref{drn}.
}
\label{ba}
\end{figure}
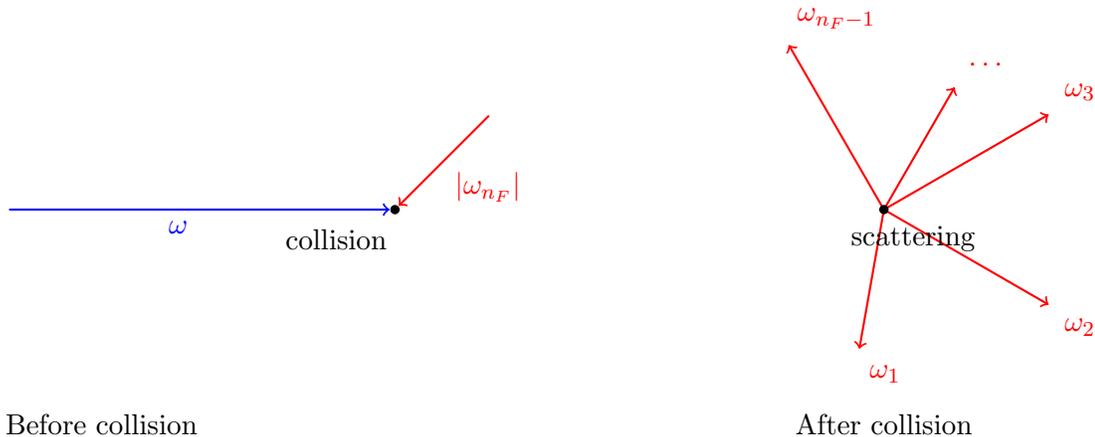

In the following, instead of considering the scattering between the particle and the Hawking radiation, we analyze the particle’s decay at finite temperature.
To compute the decay rate at finite temperature, we use \cite{Weldon:1983jn}. Essentially, it amounts to modifying the zero-temperature decay calculation as follows:\footnote{
We will ignore the inverse decay process
because we only study whether the decay is negligible or not.
}
We allow negative $\omega_f$ for final particles, which means that they represent a thermal excitation scattered by the original particle.
(see Figure \ref{ba}).
Also, we multiply a statistical factor $1+\gamma(\omega_f)$ and $\gamma(\omega_f)$ for each ``final" particle with positive and negative energy $\omega_f$, respectively, where $\gamma (\omega_f)=1/(\exp (|\omega_f|/T_{eff})-1)$.\footnote{Here, we only consider bosonic particles, but the discussions can be easily repeated for the fermionic particles. The conclusions are the same as the approximation taken in this paper.}
With these modifications, the decay rate at finite tmpereture will be given by
\begin{align}
    \frac{d \Gamma}{d \tilde{t}} = \frac{1}{2 \omega} 
    \prod_{f=1}^{n_F} 
    \left( \frac{d^3 p_f}{(2 \pi)^3} \frac{1}{2 |\omega_f|} \gamma  (\omega_f)\right) (2 \pi)^4 \delta^4 (P-\sum_{f=1}^{n_F} P_f) |{\cal A}|^2,
    \label{drn}
\end{align}
which we will qualitatively evaluate below.

The total decay rate will be 
summation of the decay rate $\Gamma$ in \eqref{drn} over all aloud ``final" states.
Here, the number of the negative frequency mode in the ``final" states is taken to be one 
because what we want to know is the scattering with the outgoing Hawking radiation rather than the thermal equilibrium.

\subsubsection*{Contributions from $|\omega_f| \gg T_{\rm eff}$ or $|\omega_f| \ll T_{\rm eff} $}

As a preparation for the decay rate calculation, we estimate how the factor $\left( \frac{d^3 p_f}{(2 \pi)^3} \frac{1}{2 |\omega_f|} \gamma  (\omega_f)\right)$ in \eqref{drn} behaves in the limits of very high and very low frequencies.
For $|\omega_f| \gg T_{eff}$, $\gamma$ becomes exponentially small as $\gamma \simeq \exp (-|\omega_f| /T_{eff})$.
Then, the statistical factor is $\gamma+1 \simeq 1$ and $\gamma \simeq 0$ for the positive and negative energy states, respectively.
Then, the negative energy ``final" state can be negligible for this case.
For $|\omega_f| \ll |T_{eff}|$,
we find $\gamma \simeq \frac{T_{eff}}{\omega_f}$which is large. However, the factor $\left( \frac{d^3 p_f}{(2 \pi)^3} \frac{1}{2 |\omega_f|}\right)$ in the decay rate \eqref{drn} gives a small factor ${\cal O}(\omega_f^2)$ after integrating out $d^3 p_f$ for the region  $|\omega_f| \ll |T_{eff}|$. Then, this IR region of integrals will be negligible
for both positive and negative energy final states. 

Below, we will estimate the decay rate for two cases: 
high-frequency mode: $\omega \gg T_{\rm eff}$
and low-frequency mode: $\omega \ll T_{\rm eff}$.

\subsubsection{High-frequency mode: $\omega \gg T_{\rm eff}$}

Now we will concentrate on evaluating \eqref{drn} for the high-frequency wavepacket $\omega_0 \gg 1/r_h$, which means $\omega \gg T_{eff}$.
Later, we will consider the low-frequency case.
We also estimate only the qualitative features of the process.
This implies that the scattering should be almost the fragmentation process \eqref{frag} of the original particle with $(\omega, {\bf p})$
because the thermal excitations contributing to the scattering occur at low energies compared with $\omega$.
Then, the final states consist of 
$n_0$ particles in the fragmentation process with
\begin{align}
    (\omega_f, {\bf p}_f)  \simeq (\alpha_f \, \omega, \alpha_f  {\bf p} ),
    \label{frag2}
\end{align}
where $\alpha_f \geq 0$ and $\sum_{f=1}^{n_0} \alpha_f=1$ for $f=1, \cdots, n_0$,
and an additional $n_F-n_0$ 
``final" states with energy $|\omega_f| \sim T_{eff}$ for $f=n_0+1, \cdots,n_F-1$. 
Here, $n_F$ is the number of the ``final" states and 
$n_0$ is the number of the ``final" states with energies comparable with $\omega$.
Here, we can approximate the statistical factor $\gamma^f$ as one for any $f$. 

Now, we will evaluate $\prod_{f=1}^{n_F} 
    \left( \frac{d^3 p_f}{(2 \pi)^3} \frac{1}{2 |\omega_f|} \gamma^f\right) (2 \pi)^4 \delta^4 (P-\sum_{f=1}^{n_F} P_f)$ in \eqref{drn}.\footnote{
We will evaluate $|{\cal A}|^2$ separately
although it depend on ${\bf p}_f$.
We can justify this by only quantitatively investigating the dominant contributions to the decay rate. 
}    
For this purpose, only the quantities we should consider in our rough approximation are $\omega$ and $T_{eff}$.
Thus, using the dimensional analysis, we will compute only the $\omega$ dependence below.

For the integrals in the decay rate \eqref{drn},
the integrals $d^3 p_f$ will be approximated as $(T_{eff})^3$ for the particle with $|\omega_f| \sim T_{eff}$.
Here, we only consider the dominant parts of the integrals, and the delta function $\delta^4 (P-\sum_{f=1}^{n_F} P_f)$ is imposed on the integrals for
the particles in the fragmentation process \eqref{frag2}.\footnote{
For $n_0=1$, another integral is needed to impose the delta function. We can easily see that the results below are correct for this case.}
For the particles of \eqref{frag2},  we approximate $\omega_f \sim \omega$.
Then, $\frac{1}{2 |\omega_f|}$ in the decay rate \eqref{drn}  give a factor $\omega^{-n_0}$ because there are $n_0$ such particles. 
On the other hand, 
the integrals $\int d^3 p_f$ for the particles contain a factor $\omega^{n_0-1}$ which comes from the momentum integrations corresponding to the variations of $\alpha_f$ in \eqref{frag2} where the number of such integrals is $n_0-1$
because of the constraint $\sum_{f=1}^{n_0} \alpha_f=1$. 
Then, we find schematically
\begin{align}
    \int \prod_{f=1}^{n_0} 
    \frac{d^3 p_f}{(2 \pi)^3} \delta^4 (P-\sum_{f=1}^{n_F} P_f)
     \sim \omega^{n_0-1} (T_{eff})^{2 n_0 -3},
    \label{dr2}
\end{align}
where $T_{eff}$ dependence was determined by the dimensional analysis, and by collecting all factors we find
\begin{align}
    \frac{d \Gamma}{d \tilde{t}} \sim \frac{1}{ \omega^2} 
     (T_{eff})^{2 n_F-3}
     |{\cal A}|^2.
    \label{dr2b}
\end{align}
Note that we need $n_F \geq 3$ for kinematical reasons.


We will evaluate the amplitude ${\cal A}$ for the decay of a bosonic field $h$. 
A general interaction term of two bosonic fields $h$ and $\phi$ in a Lagrangian\footnote{
For the fermionic fields $h, \phi$, we can discuss the same ways and show \eqref{logp} is correct.
} is schematically given by
\begin{align}
    \lambda \, (l_p)^{n_F+m-3} \, h \, \partial^m \phi^{n_F},
    \label{gl}
\end{align}
where $\lambda$ is a dimensionless coupling constant and
$h, \phi$ can have Lorentz indices, which
are contracted appropriately.
Here, $\partial^m \phi^{n_F}$ denotes all combinations with a fixed number of derivatives and fields. For example, $\partial_\mu \phi \partial^\mu \phi$ and $\phi \partial^2 \phi $ for $m=n_F=2$.
Although $h$ and $\phi$ are denoted differently for convenience, they may refer to the same field. Specifically, $h$ is used to represent the initial particle, while $\phi$ denotes the final particles.
The fields $h$ and $\phi$ are normalized such that they have the canonical kinetic term, i.e., $\frac12 \partial_\mu \phi \partial^\mu \phi$, then their mass dimension is one.
We neglect total divergence terms to obtain the form \eqref{gl}.

Of course, we need to consider the interaction including several fields $\phi_a$ for general interactions
and $\partial^m \phi^n$ can be generalized to the $n$-th product of different fields with $m$ derivatives. 
In the following, we will consider the single $\phi$ case \eqref{gl} because the analysis below is essentially the same for the generalized setup.

For the derivatives, we will approximate $\partial^2 \rightarrow \omega T_{eff} $ for $\omega_0 \gg 1/r_h$
as explained below.
As explained before, only the four vectors that are relevant for us here are the ones very close to $P=(\omega,{\bf p})$, which satisfies $P^2 \simeq 0$, and the ones with $\omega_f \sim T_{eff}$.
Then, the Lorentz invariants constructed from these 
will be $P \cdot P_{T_{eff}} \sim \omega T_{eff}$
or $P_{T_{eff}}\cdot P'_{T_{eff}} \sim (T_{eff})^2$,
where $P_{T_{eff}}$ and $P'_{T_{eff}}$ are the four momentum for the particles with $\omega_f \sim T_{eff}$.
This means that $\partial^2$ will be bounded by ${\cal O} (\omega T_{eff} )$.\footnote{
We have the polarization vector or similar quantities if we consider the spinors or the tensors.
However, in the decay rate, these are traced, and only the momentum four vectors appear.
}

Thus, up to a numerical constant, we can estimate the 
scattering amplitude coming from this interaction term as
\begin{align}
    {\cal A} \sim \lambda (l_p)^{n_F+m-3} (\omega T_{eff})^{\frac{m}{2}}(1+\cdots),
\end{align}
where $\cdots$ represent the lower order terms in $T_{eff}/\omega$.
Then, the decay rate \eqref{dr} is estimated as
\begin{align}
    \frac{d \Gamma}{d \tilde{t}} \sim 
     \frac{1}{ \omega^2} 
     (T_{eff})^{2 n_F-3} 
     |\lambda|^2 (l_p)^{2n_F+2m-6} 
     (\omega T_{eff})^{m}
     \sim 
       |\lambda|^2 (l_p)^{2n_F+2m-6} \omega^{m-2} (T_{eff})^{2n_F+m-3} \nonumber \\
       \sim   
       |\lambda|^2 \left( \frac{x}{l_p} \right)^{-2n_F-2m+5} \frac{1}{l_p}(r_h \omega_0 )^{m-2}.
       \label{decay}
\end{align}

Now, we will evaluate the probability such that the wave packet does not decay when the wave packet is on $x \in [r_h, e^\beta l_p]$, where $\beta$ is a ${\cal O}(1)$ parameter and we will set 
$\beta=0$ when we think the particle comes to the "location" of the stretched horizon.
We can clearly ignore the quantum gravity effects when it is on $x > r_h$, which means $r \gtrapprox 2 r_h$, because the blueshift effect is not large.
This probability is given by
\begin{align}
    P=\exp\left( {-\sum_a \int_{e^\beta \, l_p}^{r_h} dx \frac{d \Gamma_a}{dx}} \right),
    \label{int}
\end{align}
where $a$ takes all the decay modes and 
$\Gamma_a$ is the corresponding decay rate.
We apply the formula \eqref{decay} for the small interval 
and the energy $\omega$ should be 
regarded as effective energy $\omega(\bar{x})$ at $x=\bar{x}$.
Then, the part of $\log P$ for the interaction \eqref{gl} is given by
\begin{align}
    -\int_{e^\beta \, l_p}^{r_h} dx \frac{d \Gamma}{dx} \simeq - c \int_{e^\beta \, l_p}^{r_h} dx 
       |\lambda|^2 \left( \frac{x}{l_p} \right)^{-2 \Delta-1} \frac{1}{l_p}(r_h \omega_0 )^{m-2} 
    = - \frac{c}{\Delta \, e^{2 \Delta \beta} }
    |\lambda|^2
     (r_h \omega_0)^{m-2}
    +{\cal O}((l_p)^{2 \Delta})
    \label{logp}
\end{align}
where $c$ is a numerical constant of the order of unity, $\Delta=n_F+m-3$, which is the mass dimension of the coupling constant of the interaction term \eqref{gl}. 
Here, we assumed $\omega_0 \gg 1/r_h$ and $\Delta >0$.

For $\Delta  < 0$, the integral of \eqref{logp} 
is dominant for $x \gg l_p$ and gives a large value if the $\lambda$ is not sufficiently small.
Thus, $\lambda$ is assumed to be sufficiently small for $\Delta <0$
because the decay rate should be very small
for the region where the blueshift effect is not very large,
as discussed later again.
In other words, if $\lambda$ is not sufficiently small, the quantum gravity effect is manifest even in the asymptotic flat space region, which is not the theory we want to consider.
For $\Delta =0$, \eqref{logp} will be replaced by 
\begin{align}
   -\int_{e^\beta \, l_p}^{r_h} dx \frac{d \Gamma}{dx} \simeq - c
    |\lambda|^2
     (r_h \omega_0)^{m-2} 
     \log \left( \frac{r_h}{e^\beta \, l_p} \right),
    \label{logp0}
\end{align}
which depends on $\log (l_p/r_h)$ and can take very large values.
However, for the marginal coupling $\Delta=0$,
the number of derivatives will be $1$ or $0$
and $\lambda$ will not be large near the Plank scale.
Thus, this term will be rather small and not important compared with $\Delta >0 $ contributions.

It is important to note that the leading term of \eqref{logp}
is $l_p$ independent (if we assume $\lambda$ is $l_p$ independent).  
Furthermore, we can easily check that the (leading order) decay rate is 
$l_p$ independent even if we drop the assumption, $\omega_0 \gg 1/r_h$.
This can be explained intuitively as follows.
The integral in \eqref{logp}
is dominant for $x = {\cal O}(l_p)$,
i.e. near the stretched horizon.
In the region, all the quantities of mass dimension one we considered here are
proportional to $1/l_p$, except the upper limit of the integral in \eqref{logp}, because of the blueshift.
Here, we assumed $\lambda$ is $l_p$ independent.
Because the probability $P$ is dimensionless, it should be independent of $l_p$.

\subsubsection{Small-frequency mode: $\omega \ll T_{\rm eff}$}
Here, we will consider the $\omega_0 \ll 1 /r_h$ case.
The scattering of this particle and a particle of the Hawking radiation with four momentum $P_{T_{eff}}$ is kinematically equivalent to case $\omega_0 \gg 1 /r_h$ we considered earlier, with the roles of the two particles interchanged. As a result, the outgoing particles can be approximately viewed as a fragmentation of $P_{T_{eff}}= (\omega_{T_{eff}}, {\bf p}_{T_{eff}})$, accompanied by particles with energy of order $\omega$.
Then, the final states consist of 
$n_0$ particles in the fragmentation process with
\begin{align}
    (\omega_f, {\bf p}_f)  \simeq (\alpha_f \, \omega_{T_{eff}}, \alpha_f  {\bf p}_{T_{eff}} ),
    \label{fragT}
\end{align}
where $\alpha_f \geq 0$ and $\sum_{f=1}^{n_0} \alpha_f=1$ for $f=1, \cdots, n_0$,
and additional $n_F-n_0-1$ final states with energy $\omega_f \sim \omega$ for $f=n_0+1, \cdots,n_F-1$.
Note that the "final" state of $f=n_F$ should have 
a negative $\omega_f \sim -T_{eff}$ \cite{Weldon:1983jn}, which represent the initial particle of the Hawking radiation with  $P_{T_{eff}}$.
Here, $n_F$ is the number of the ``final" states and 
$n_0+1$ is the number of the ``final" states with energies comparable with $T_{eff}$.

Now, we will evaluate $\prod_{f=1}^{n_F} 
    \left( \frac{d^3 p_f}{(2 \pi)^3} \frac{1}{2 |\omega_f|} \gamma (\omega_f)\right) (2 \pi)^4 \delta^4 (P-\sum_{f=1}^{n_F} P_f)$ in \eqref{drn} as before.
For the integrals in the decay rate \eqref{drn},
the integrals $d^3 p_f$ will be approximated as $\omega^3$ for the particle with $\omega_f \sim \omega$.
Here, we only consider the dominant parts of the integrals, and the delta function $\delta^4 (P-\sum_{f=1}^{n_F} P_f)$ is imposed on the integrals for
the particles in the fragmentation process \eqref{fragT}.
Then, we find $\int \prod_{f=n_0+1}^{n_F-1} 
    \left( \frac{d^3 p_f}{(2 \pi)^3} \frac{1}{2 |\omega_f|} \right) \sim \omega^{2(n_F-n_0-1)}$.
For the particles of \eqref{fragT},  we approximate $\omega_f \sim T_{eff}$.
Then, following the arguments for $\omega \gg T_{eff}$,
we find
\begin{align}
    \int \prod_{f=1}^{n_0} 
    \frac{d^3 p_f}{(2 \pi)^3} \delta^4 (P-\sum_{f=1}^{n_F} P_f)
     \sim (T_{eff})^{n_0-1} (\omega)^{2 n_0 -3},
    \label{drT1}
\end{align}
and 
\begin{align}
    \frac{d \Gamma}{d \tilde{t}} \sim  T_{eff} 
     (\omega)^{2 n_F-6}
     |{\cal A}|^2,
    \label{drT2}
\end{align}
by collecting all factors.

The scattering amplitude will be approximated as before:
\begin{align}
    {\cal A} \sim \lambda (l_p)^{n_F+m-3} (\omega T_{eff})^{\frac{m}{2}},
\end{align}
where we neglected smaller terms by $\omega/T_{eff}$.
Thus, the decay of this process is given by 
\begin{align}
    \Gamma= c \int_{e^\beta \, l_p}^{r_h} dx 
       |\lambda|^2 \left( \frac{x}{l_p} \right)^{-2 \Delta-1} \frac{1}{l_p}(r_h \omega_0 )^{2 n_F-6+\frac{m}{2}} +\cdots
    = - \frac{c}{\Delta \, e^{2 \Delta \beta} }
    |\lambda|^2
     (r_h \omega_0)^{2 n_F-6+\frac{m}{2}}
    +\cdots,
    \label{logp2}
\end{align}
where $c$ is a order $1$ numerical constant,
\begin{align}
\Delta=n_F+m-3,    
\end{align}
which is the mass dimension of the interaction term \eqref{gl} and we assume $\omega_0 \ll 1/r_h$. 
Here, $\cdots$ denotes the terms of order ${\cal O}((r_h \omega_0)^{2 n_F-5+\frac{m}{2}})$ and 
the Planck suppressed terms, and
we assumed $\omega_0 \ll 1/r_h$ and $\Delta >0$.
For $\Delta  \leq  0$, the same argument can be made as in the case $\omega_0 \gg 1/r_h$.

The largest contributions to the decay rate will be 
given by the interactions with $n_F=3$ and $m=0$, i.e., the four-point interactions with no derivatives.
Note that the interaction with $n_F < 3$ can not give the decay because of the kinematical constraint for the massless particles.
However, for the graviton, we will see that any four-point interaction without derivatives
may have a coupling constant $\lambda$ which is significantly suppressed by $l_p/r_h$.
Thus, the main contribution without the Planckian suppression in the coupling constant comes from interactions $(l_p)^2  h^2 (\partial h)^2 \subset \frac{1}{G_N} \sqrt{-g} {\cal R}$, which corresponds to $n_F=3$ and $m=2$ in \eqref{gl}. Here, $g$ is the determinant of the metric and ${\cal R}$ is the Einstein-Hilbert action. Finally, we obtain the leading order of the decay rate as
\begin{align}
    \Gamma= - \frac{c}{\Delta \, e^{2 \Delta \beta} }
    |\lambda|^2
     (r_h \omega_0)
    +\cdots.
    \label{logp2l}
\end{align}

It should be noted that the decay probability \eqref{int} contains 
the infinite summation over the various final states.
The infinite summations are the summation over the number of the derivatives, $m$, and over the number of the final state, $n_F$.
These infinite sums could potentially lead to issues like divergence, but we argue that this is not the case, as is explained below. We evaluate the numerical constants that we have previously neglected. First, for high frequency case, the sum over \( m \) rapidly decreases as \( m \) becomes large. Additionally, since the incident particle scatters with the Hawking radiation particle, the sum of these two particles is distributed among the final particles. 
This introduces a factor of approximately \( \frac{1}{n_F - 1} \) for the energy of each final particle.\footnote{Such an equally distributed energy will give the largest contribution of the momentum integrals. We also note that, to be more precise, the energies are approximately \( \frac{\omega}{n_0} \) and \( \frac{T_{\text{eff}}}{n_F - n_0 - 2} \).} 
As a result, the sum over \( n_F \) becomes very small as \( n_F \) increases.
For low frequencies case, 
the energies of the final particles are roughly $\frac{T_{eff}}{n_F-1}$ and
then the decay rate has a factor
$\frac{1}{(n_F-1)^{2 n_F-4}}$.
This factor also gives $\frac{1}{(n_F-1)^m}$ for the decay rate.
Thus, the contributions with small $m$ and $n_F$ may approximate the infinite sum.

\section{Estimation of the coupling for the real world}

Here, we will consider the interaction terms in the real world and estimate their coupling constants.
The (effective) action will be the Einstein-Hilbert action plus the phenomenological model of the elementary particles
with 
Planck-suppressed higher derivative terms. 
Here, the decay we consider will occur almost at the Planck scale.
Thus, we will consider the effective action near this scale.

We will parametrize the dimensionless coupling $\lambda$ in \eqref{gl} as $\lambda=(l_p M)^{-\Delta}$ where $M$ is a parameter of mass dimension one for $\Delta \neq 0$.

For a relevant coupling $(\Delta<0)$, which includes the cosmological constant and the mass term,
$M$ should satisfy $M \ll 1/l_p$, which means $\lambda \ll 1$, because there are no coupling with $M \gtrsim 1/l_p$ in the low energy theory by the assumption.
Thus, such coupling gives a negligible decay mode, at least compared with other decay modes.
For an irrelevant coupling $(\Delta>0)$, 
$M$ is expected to satisfy $M \gtrsim 1/l_p$, which means $\lambda \lesssim 1$.
Thus, only the terms with $\Delta \geq 0$ are important and will have a coupling $\lambda ={\cal O}(1)$.
For example, 
the Lagrangian for the Einstein-Hilbert action $\frac{1}{G_N} \sqrt{-g} \cal R$ contains the term $(l_p)^2 h^2 \partial h \partial h$ which is $m=2$ and $n_F=3$
and the effective action may contain $\sqrt{-g} {\cal R}^2$ term, including terms of $m=4$.
Note that $\cal R$ may contain the term $h^4$ for which the derivatives act on the (non-trivial) background metric,
but the coupling constant $\lambda$ for this term is ${\cal O}((l_p/r_h)^2)$.
If there is a (scalar) matter field $\phi$,
the effective action can contain the term $\int d^4 x \, \sqrt{-g} \lambda  \phi^4$, whose expansion  
contains the term $\lambda h^2 \phi^4$ where $h$ is the graviton field.
This is a $m=0$ term.

Then, 
the decay probability for $\omega_0 r_h \gg 1$ is almost zero, i.e. $P \approx 0$ because of \eqref{logp} with $m >2$. 

For $\omega_0 r_h \ll 1$, 
the decay probability, which is determined by 
the \eqref{logp2}, will be 
\begin{align}
    \log P=- \sum_{n=0}^\infty  c_n (r_h \omega_0)^{n+1},
    \label{cn}
\end{align}
with  numerical coefficients $c_n$, which satisfy $c_n \geq 0$ and $c_n= {\cal O} ((\omega_0 r_h)^0)$ for the graviton.
Here, we include subleading terms for $\omega_0 r_h \ll 1$ and assume that this summation is convergent for a sensible theory.


\section{Application to an astrophysical black hole}

Let us consider an astrophysical black hole with a typical size.
First, we will consider a particle with a wavelength that is much less than the black hole size, i.e. 
$(r_h \omega_0 \gg 1)$.
For example, the photons of radio waves or even high-energy radiation satisfy this condition.
In that case, as expected, we find the probability $P \approx 0$ and
the particle almost completely decays before it reaches the stretched horizon.
The particles produced by the decay will decay again until they become low-energy particles near the stretched horizon.
This is regarded as a kind of thermalization, and they are finally emitted from the black hole, regarded as Hawking radiation.
Thus, this is consistent with the observations and semi-classical picture of the black hole, which is ``black" at the high-frequency regime.

For a particle with a wavelength that is comparable with the black hole size, i.e. 
$(r_h \omega_0 \lesssim 1)$,
we find $P ={\cal O}(1)$.
Thus, if we assume a (partial) reflection of incoming waves at the stretched horizon (or the brick wall),
we may observe the reflected outgoing waves.\footnote{
It may be possible that there is no reflection and complete absorption at the stretched horizon,
but we expect that it is not likely.
This is because the degree of freedom will be exhausted outside the stretched horizon, as discussed in \cite{Iizuka:2013kma} \cite{Terashima:2021klf} in the AdS/CFT context, and then the dissipation at the stretched horizon may not be possible and some part of the wave may reflect.
}

Now, let us consider the gravitational waves produced by a merger of a binary compact object, such as a black hole-black hole or black hole-neutron star binary, leading to the emission of black hole ringdown after the merger phase.
Some produced gravitons or other species will fall into the black hole.
We can observe the (possible) reflected waves with 
$(r_h \omega_0 \lesssim 1)$ suppressed by the decay with an energy-dependent factor $ e^{-c_0 (r_h \omega_0) -c_1 (r_h \omega_0)^2-\ldots}$ and we find
\begin{equation}
P (\omega) = P_0 \,\, \exp({-c_0 (r_h \omega_0) -c_1 (r_h \omega_0)^2-\ldots}),
\end{equation}
where $P_0$ is the reflection probability at or the inside the stretched horizon, for example, for the brick wall model $P_0=1$.




What we would like to emphasize here is that our result estimates the suppression of the reflection probability due to effects near the stretched horizon, independently of how the quantum black hole is traversed (as characterized by \( P_0 \)). In particular, in the case where perfect reflection is assumed, i.e., \( P_0 = 1 \), 
the leading exponent in the formula is given by
\begin{equation}
P (\omega) = \exp{(c_0 (r_h \omega)}),
\end{equation}
and this is already known as the Boltzmann-reflectivity model \cite{Oshita:2018fqu}, where the authors phenomenologically introduced a dissipation term in the perturbation equation of a black hole. The dissipation term is blue-shifted near the horizon. It leads to the reflectivity of $e^{- c \omega / T_{\rm H}}$ where $c$ is a constant parameter and $T_{\rm H} \sim {\cal O}(1/r_h)$ is the Hawking temperature. In this work, we find that our formula agrees with the Boltzmann reflectivity at the leading level in the low-frequency regime. Based on this, we conclude that the Boltzmann model is a reasonable model even from the microscopic point of view.
Thus, for $P_0=1$, a numerical computation of GW waveform sourced by a particle plunging into the black hole with the reflectivity of $P(\omega)$ will be similar to the one in \cite{Oshita:2018fqu} although we have not done it.

\section{Comments on the assumptions we made}

So far, several assumptions have been made implicitly at times. Here, we will clarify some of the assumptions and discuss to what extent they can be justified.

The first issue is that
we are interested in 
wave packets and also semi-classical waves falling into the black hole.
Assuming the 
semi-classical waves can be described by a weakly interacting quantum field theory on the fixed background,
they may be considered as coherent states,
which are linear combinations of multi-particle states.
Then, we can apply the decay rate to each particle.  
The probability $P$ of each particle may be related to the rate of decrease of energy flux.

The second issue is that the scattering cross-section is calculated for a long-duration uniform beam, but we applied it to a short-duration case. This was because the scattering intensity changes as it approaches the horizon. Although we have only performed qualitative calculations, this could still become an issue near the stretched horizon. A related problem is that when considering a wave packet, its size needs to be larger than the inverse of the frequency, and the scattering intensity is not constant on the wave packet.
Here, what was important in our calculation is that \(x\) is of the order of \(l_p\), and in that region, the frequency of the wave packet is also of the same order. Therefore, while \(\mathcal{O}(1)\) corrections are naturally possible, our calculation is expected to be qualitatively correct.

The third issue is that our scattering calculation assumes plane waves, whereas, in the case of phenomena such as black hole echoes, the waves are spherical. (For high-frequency cases involving wave packets, the plane wave approximation is considered reasonable, as in typical particle scattering experiments.)
This difference might be important, but we expect it is not. This is because the black hole radius is huge compared with the Planck length and
the angular momentum of the one particle of the coherent state will be tiny.
Then, the approximation of the spherical wave to the plane wave may not be bad qualitatively.





\section*{Acknowledgement}

The author would like to thank N. Oshita for collaboration in the early stages of this work
and helpful discussions and comments.
The author would like to thank T. Kawamoto and S. Sugishita for the useful comments.
This work was supported by MEXT-JSPS Grant-in-Aid for Transformative Research Areas (A) ``Extreme Universe'', No. 21H05184.
This work was supported by JSPS KAKENHI Grant Number 	24K07048.

\hspace{1cm}



\bibliographystyle{utphys}
\bibliography{main202409.bib}
\end{document}